\documentclass[conference]{IEEEtran}

\usepackage{graphicx}  

\begin{document}

\title{Domain wall switching: optimizing the energy landscape}

\author{\authorblockN{Zhihong Lu}
\authorblockA{MINT Center and Department of Physics and Astronomy\\
University of Alabama, Tuscaloosa AL 35487-0209\\
Email: zlu@mint.ua.edu}
\authorblockN{P. B. Visscher}
\authorblockA{MINT Center and Department of Physics and Astronomy\\
University of Alabama, Tuscaloosa AL 35487-0209\\
Email: visscher@ua.edu} 
\authorblockN{W. H. Butler}
\authorblockA{MINT Center and Department of Physics and Astronomy\\
University of Alabama, Tuscaloosa AL 35487-0209\\
Email: wbutler@mint.ua.edu} }


\maketitle

\begin{abstract}
It has recently been suggested that exchange spring media offer a
way to increase media density without causing thermal instability
(superparamagnetism), by using a hard and a soft layer coupled by
exchange.  Victora has suggested a figure of merit %
$\xi = 2 E_b/\mu_0 m_s H_{sw}$, the ratio of the energy barrier to
that of a Stoner-Wohlfarth system with the same switching field,
which is 1 for a Stoner-Wohlfarth (coherently switching) particle
and 2 for an optimal two-layer composite medium. A number of
theoretical approaches have been used for this problem (e.g.,
various numbers of coupled Stoner-Wohlfarth layers and continuum
micromagnetics).  In this paper we show that many of these
approaches can be regarded as special cases or approximations to a
variational formulation of the problem, in which the energy is
minimized for fixed magnetization. The results can be easily
visualized in terms of a plot of the energy $E$ as a function of
magnetic moment $m_z$, in which both the switching field [the
maximum slope of $E(m_z)$] and the stability (determined by the
energy barrier $\Delta E$) are geometrically visible.  In this
formulation we can prove a rigorous limit on the figure of merit
$\xi$, which can be no higher than 4.  We also show that a quadratic
anistropy suggested by Suess \textit{et al} comes very close to this
limit.
\end{abstract}

\IEEEpeerreviewmaketitle

\section{Introduction}

Recently the concept of an exchange-spring
medium\cite{victora,wang,dobbin} whose grains have a soft and a hard
layer has been generalized\cite{suess&} to a system with a
continuously-varying anisotropy.  Various models of this system have
been explored -- the purpose of this paper is to show that the
relationships between these can be easily visualized by using a
variational formulation of the problem.

In an exchange-spring medium, we want to minimize the switching
field, but of course we can make this as small as we want by using a
very small anisotropy, and the medium will be superparamagnetic
(thermally unstable) and useless.  To make comparisons between
media, we must hold something constant to maintain stability.  Often
what is held constant is the anisotropy field (the coercivity
$2K/M_s$) of the hardest layer. However, this hardest layer might be
very thin and have little effect on the overall coercivity.  In a
practical application the quantity it is most important to hold
constant is the overall energy barrier to switching, which we will
assume determines the thermal stability of the medium.
Victora\cite{victora} introduced a figure of merit for this purpose, %
$\xi = 2 E_b/\mu_0 m_s H_{sw}$ (here $E_b$ and $m_s$ are the barrier
energy and saturation magnetic moment per unit area) for which we
will prove a rigorous bound in Sec. \ref{rigor}.

To do this, we develop a variational formulation in which we
describe the switching behavior in terms of a function $E(m_z)$, the
energy per unit area as a function of the magnetic moment per unit
area. This turns out to be a very useful way of thinking about
switching problems.

\section{Model}
We consider a one-dimensional model, in which the magnetization
$\mathbf{M}(z)$ is a function only of one variable $z$ (independent
of $x$ and $y$).   We will allow the anisotropy $K(z)$, exchange
constant $A(z)$, and saturation magnetization $M_s(z)$ to vary
arbitrarily with $z$. Since we will do computations with a discrete
approximation to this continuum model (which approaches the
continuum model as the cell size $\rightarrow 0$), we will write the
energy in a discrete form.  It has cells labeled by $i$, with
magnetization vectors $\mathbf{M}_i$.  In the quasistatic energy
minima we will consider, these vectors will lie in a plane, so they
can be described by giving the angle $\theta_i$ of the magnetization
relative to the long axis of the grain (the $z$ axis):

The energy (per unit area in the $xy$ plane) $E$ of our system is
then given in terms of the values of $K$ and $M$ at each cell (and
$A$ between each neighboring pair of cells) by
\begin{eqnarray}
E = \sum_{i=1}^N a_i K_i \sin^2 \theta_i + %
\sum_{i=1}^{N-1} a_i \frac{2A_{i,i+1}}{a_{i,i+1}^2}
\cos(\theta_{i+1}-\cos\theta_i) + \nonumber \\
\sum_{i=1}^N a_i \mu_0 M_i H \cos \theta_i   \label{energy}
\end{eqnarray}%
where $K_i$ is the (perpendicular) anisotropy at the center of cell
$i$, $a_i$ is the length of cell i, $a_{i,i+1}$ is the distance
between cells $i$ and $i+1$, $A_{i,i+1}$ is the continuum exchange
parameter evaluated between these cells\cite{aha}, and $H$ is the
external field (assumed along $z$).  For simplicity, we do not
consider magnetostatic energy here -- in similar systems,
micromagnetic simulation has shown that this affects the coercivity
by only a few percent.

We consider here quasistatic switching –- we assume that we vary $H$
in such a way that the system is always at a relative minimum (with
respect to the $\mathbf{M}_i$'s) of the energy.   More precisely, we
assume $H$ is very slightly above this value, so that the magnetic
moment
\begin{equation}
m_z = \sum_i^N a_i M_i \cos \theta_i   \label{mz}
\end{equation}%
increases slowly, and we consider the limit in which the rate of
increase approaches zero.  Note that this is never true in a real
switching event –- after a domain wall has traversed most of the
sample, it would require reversing $H$ to keep the system
quasistatic.  However, by this time it is irrelevant whether the
system remains quasistatic (it will finish switching in either case)
and in the initial stages the quasistatic assumption is often
reasonable.

It would appear that to find the quasistatic switching trajectory,
in which $H$ varies with time, we would need to minimize a function
$E(\theta_1, \theta_2, ... \theta_N,H)$ of a large number of
variables $\theta_1, \theta_2, ... \theta_N$, for each value of $H$
independently. However, there is a way around this. We can choose
some coordinate in the space of $\theta_i$'s (we choose the
longitudinal component of the magnetic moment, $m_z$, for reasons
apparent below) and first minimize %
$E(\theta_1, \theta_2, ... \theta_N,H)$ for fixed $m_z$, obtaining a
function (the constrained minimum energy) $E(m_z,H)$. Then we can
minimize $E(m_z,H)$ with respect to $m_z$, obtaining the same
relative minimum $E(H)$ we would have obtained by unconstrained
minimization. [Note that there may be more than one relative
minimum, so we should call this $E_j(H)$ where $j$ indexes the
minima, but we will omit this index for simplicity.]  The advantage
of this apparently-circuitous method of finding the minimum is that
the configuration minimizing $E(\theta_1, \theta_2, ... \theta_N,H)$
is actually independent of $H$! This is apparent from Eq.
(\ref{energy}) above, since the only dependence on $H$ is the Zeeman
term $\mu_0 m_z H$, which is a constant when $m_z$ is held fixed.
The result is that we need only compute the constrained minimum
energy at $H=0$, and it is given at any other field $H$ by
\begin{equation}
E(m_z,H) = E(m_z,0) - \mu_0 m_z H   \label{EH}
\end{equation}%
Furthermore, this energy is minimized at a particular $H$ by setting
$\partial E(m_z,H) / \partial m_z = 0$, so the field necessary to
hold $m_z$ constant is given by
\begin{equation}
\mu_0 H = \frac{\partial E(m_z,0)}{\partial m_z}  \label{H}
\end{equation}%

\begin{figure}[tbh]
\begin{center}
\includegraphics[height=2.6 in]{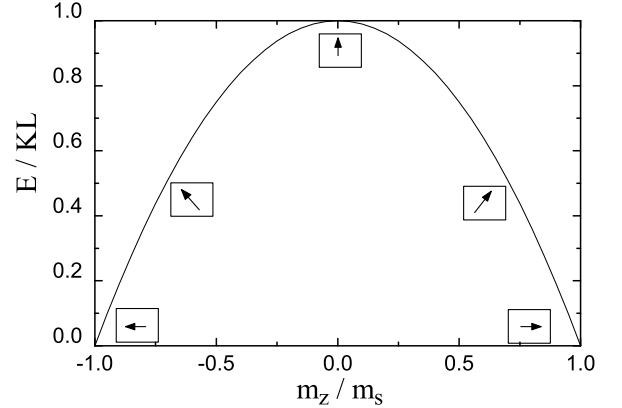}
\end{center}
\par
\vspace{-0.3 in} 
\caption{Energy landscape $E(m_z)$ for a Stoner-Wohlfarth particle.}
\label{SW}
\end{figure}

We conclude that everything we need to know about the system (the
coercivity and energy barrier) is contained in the function
$E(m_z)$, the minimum energy at fixed magnetic moment $m_z$ and zero
field. This result is very general.  Although we motivated it above
by considering domain-wall switching, it describes Stoner-Wohlfarth
(S-W) switching as well.  This is the limit in which $K$, $A$, and
$M_s$ are uniform and $A$ is large so $\textbf{M}(z)$ is uniform.
The S-W energy (per unit area, of a grain of length L) is just $E =
KL \sin^2 \theta  = KL(1-m_z^2/m_s^2)$ (here the saturation moment
per unit area is $m_s=M_s L$) so the $E(m_z)$ plot is a parabola, as
shown in Fig. \ref{SW}.
Note that the slope (the field necessary to switch, Eq. \ref{H}) is
exactly the Stoner-Wohlfarth switching field $H_{sw} = 2K/M_s$, as
expected. [In general, the coercivity is the maximum value of the
slope.]

\begin{figure}[tbh]
\begin{center}
\includegraphics[height=2.8 in]{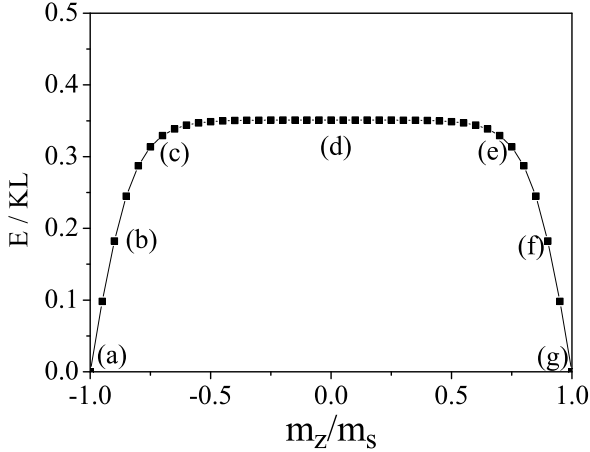} 
\end{center}
\par
\vspace{-0.3 in} 
\caption{Energy landscape $E(m_z)$ for a particle with small
exchange parameter $x=0.0076$.} \label{wall}
\end{figure}

\begin{figure}[tbh]
\begin{center}
\vspace{-0.1 in}
\includegraphics[height=2.6 in]{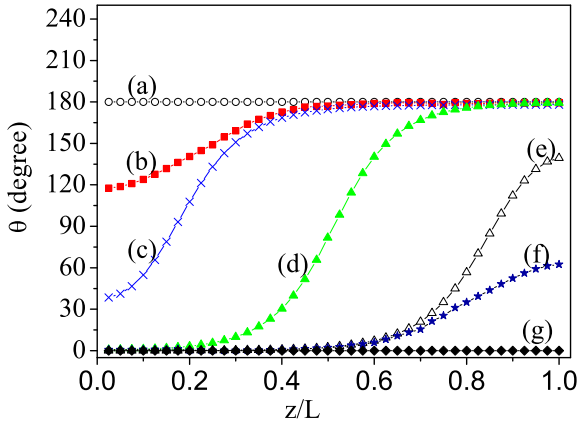} 
\end{center}
\par
\vspace{-0.3 in} 
\caption{Magnetization profiles (angle vs. position z) for the
particle whose energy landscape is shown in Fig. \ref{wall}. Labels
(a), (b), ... correspond to specific values of $m_z$ shown in Fig.
\ref{wall}.} \label{theta}
\end{figure}

Another virtue of the function $E(m_z)$ is that it has exactly the
same interpretation for a particle with a lower exchange constant
$A$ as for a S-W (high-$A$) particle. The behavior depends only on
the dimensionless parameter $x = A/KL^2$, which is the square of the
ratio of the exchange length to the particle length $L$.   With low
$x$, switching takes place through domain wall motion, and the
energy barrier is approximately the domain wall energy. This can be
calculated analytically in an infinite system (we will refer to this
as the thin-wall approximation, because it is valid when the wall is
far from the system boundary and the material properties vary only
slightly through the wall) -- the thin-wall energy is
$4(AK)^\frac{1}{2} = 4 K L x^\frac{1}{2}$.  We have developed a
numerical minimization program for computing $E(m_z)$ for an
arbitrary $K(z)$, and the result for a uniform $K(z)$ is shown in
Fig. \ref{wall}.
It can be seen that the energy is indeed constant when $m_z$ is far
from its limiting values $\pm m_s \equiv \pm M_s L$, and equal to
$4(AK)^\frac{1}{2}$.  The slope of $E(m_z)$ at the ends is just the
domain wall nucleation field. The behavior of the magnetization
profile at various times during switching is shown in Fig.
\ref{theta}. %

\begin{figure}[tbh]
\begin{center}
\includegraphics[height=2.8 in]{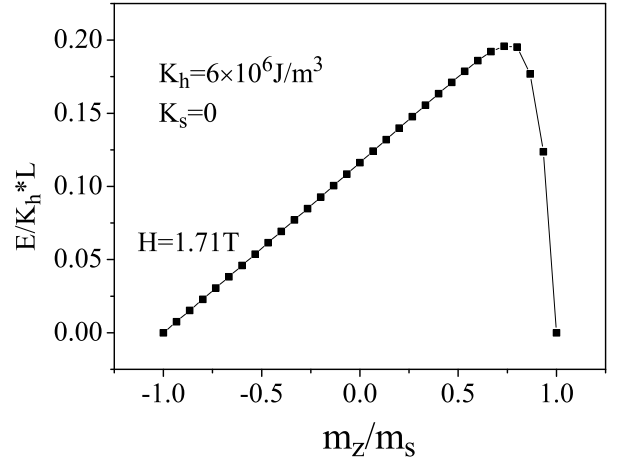} 
\end{center}
\par
\vspace{-0.3 in} 
\caption{Energy landscape $E(m_z)$ for the case $K(z)=(K_h/L^2)
z^2$, which would be exactly linear in the thin-wall approximation.
We have used $K_h=6\times 10^6 \textrm{J/m}^3$, $L=22.2$ nm,
$A=1.0\times 10^{-11}$ J/m, so that the dimensionless exchange
parameter $A/K_h L^2 = 0.00338$.} \label{z2}
\end{figure}

Suess et al\cite{suess} have noted that in the thin-wall
approximation, the pinning field should remain constant if we choose
$K(z) \propto z^2$.  In our $E(m_z)$ formulation, this means the
slope should be nearly constant. This turns out to be remarkably
nearly true numerically, except near the hard end, as shown in Fig.
\ref{z2}.

\section{Rigorous bound on coercivity figure of merit}\label{rigor}

Our $E(m_z)$ formulation allows us to prove a completely general
(within the assumptions: 1D, quasistatic) result, which is clear
geometrically from the $E(m_z)$ graph.  If we fix the vertical
height (the zero-field barrier $E_b$) and the horizontal extent
($2m_s$) the minimum possible coercivity (coercivity = maximum
slope) is obtained by a straight line, whose slope must be $\mu_0 H
= E_b / 2m_s$.  In terms of the figure of merit, this means %
$\xi \leq 4$.

Another way of stating this result is that the coercivity of any
graded medium cannot be less than 1/4 of the coercivity of a
Stoner-Wohlfarth particle (assuming that the latter switches
coherently) of the same magnetic moment and energy barrier. [If
$M_s$ is constant, fixing the moment is the same as fixing the
length $L$.] Note that the $K(z) \propto z^2$ case (Fig. \ref{z2})
gives $\xi = 3.23$, close to the theoretical limit, which it
approaches as $x \rightarrow 0$.

Note that in this paper we consider only fields along the easy axis.
Obviously it is worth considering how transverse fields might be
useful in switching, since it is known that by using a field at $45
^\circ$ from the axis the Stoner-Wohlfarth switching field is
decreased by a factor of 2, so the figure of merit $\xi$ becomes 2.
Also, it is likely that the nucleation of a domain wall (which
initially requires transverse twisting of the magnetization) can be
assisted by a transverse field, so the figure of merit might
increase slightly above 4.  We should also note that the limit we
have established assumes fixed grain length -- it may be possible to
decrease switching fields beyond the factor of 4 because graded
media may make it possible to use longer grains without encountering
complicated switching modes such as vortices.

\section{Conclusion}
We have shown a general bound on the figure of merit of a
graded-anisotropy medium, and that this bound (4) is very nearly
achieved by $K(z) \propto z^2$.  Because the usefulness of such a
medium depends on its thermal stability as well as its coercivity,
and because of the complex switching mechanism the zero-field
switching rate is not completely determined by the energy barrier,
an important remaining problem is the more precise calculation of
this rate.  Although brute force micromagnetic simulation of such
slow switching is not practical, work is under way on accelerated
sampling techniques for solving this problem\cite{bounce}.

\section*{Acknowledgment}
This work was supported by NSF MRSEC grant DMR-0213985 and by the
DOE Computational Materials Science Network.

\end{document}